\documentclass[aps,prl,10pt,twocolumn,superscriptaddress]{revtex4-1}
\usepackage{amsmath}
\usepackage{amssymb}
\usepackage{gensymb}
\usepackage{bbm}
\usepackage{graphicx}
\usepackage{framed}
\usepackage{color}
\usepackage{hyperref}
\usepackage{epstopdf}
\usepackage{dsfont}
\usepackage{amssymb}
\usepackage{amsmath}
\usepackage{amsthm}
\usepackage{wrapfig}
\usepackage{bm}
\usepackage{enumitem}
\usepackage[utf8]{inputenc} 
\usepackage[T1]{fontenc}

\usepackage{pifont}

\newcommand{\ket}[1]{\vert#1\rangle}
\newcommand{\bra}[1]{\langle#1\vert}

\newcommand{\ba}{\begin{eqnarray}}
\newcommand{\ea}{\end{eqnarray}}

\begin{document}

\title{Anomalous values, Fisher information, and contextuality, in generalized quantum measurements}  

\author{Valeria Cimini }
\affiliation{Dipartimento di Scienze, Universit\`{a} degli Studi Roma Tre, Via della Vasca Navale 84, 00146, Rome, Italy}

\author{Ilaria Gianani}\email{ilaria.gianani@uniroma3.it}
\affiliation{Dipartimento di Scienze, Universit\`{a} degli Studi Roma Tre, Via della Vasca Navale 84, 00146, Rome, Italy}
\affiliation{Dipartimento di Fisica, Sapienza Universit\`{a} di Roma, P.le Aldo Moro, 5, 00185, Rome, Italy}

\author{Fabrizio Piacentini}
\author{Ivo P. Degiovanni}
\affiliation{INRIM, Strada delle Cacce 91, 10135, Torino, Italy.}

\author{Marco Barbieri}
\affiliation{Dipartimento di Scienze, Universit\`{a} degli Studi Roma Tre, Via della Vasca Navale 84, 00146, Rome, Italy}
\affiliation{Istituto Nazionale di Ottica - CNR, Largo Enrico Fermi 6, 50125, Florence, Italy}

\begin{abstract} 

Postselection following weak measurements has long been investigated for its peculiar manifestation of quantum signatures. In particular, the postselected events can give rise to anomalous values lying outside the spectrum of the measured quantity, and may provide enhanced Fisher information. Furthermore, the Pusey inequality highlights that, for extremely weak measurements, non-contextual models can account for the outcome probabilities. It is then interesting to investigate whether these are linked in a unified framework. Here we discuss on the existence of a possible connection in the case of qubits. We show that when performing generic postselected measurements there exist no one-to-one mapping between them, an instance that leads to drawing more involved considerations.

\end{abstract}

\maketitle

{\it Introduction.} The outcome of a measurement carried out on a spin-1/2 particle can turn out to be 100~\cite{PhysRevLett.60.1351,PhysRevLett.94.220405}. What is exactly measured in these procedures, and whether this carries physical meaning has been debated ever since the introduction of the concept of postselected values~\cite{PhysRevLett.62.2325,PhysRevLett.62.2326,PhysRevD.40.2112,doi:10.1098/rsta.2016.0395,PhysRevA.96.032114}. These anomalies emerge when an observable is measured through an indirect procedure, i.e. by inferring its value by the coupling of the spin to an auxiliary probe system and only accessing the latter.

The state of this second system needs not being optimized to deliver full information at each shot, since the expectation value can be recovered exactly from a large collection of events. Each individual event does not provide unambiguous information on the observable ~\cite{RevModPhys.86.307}, hence resulting in proportionally reduced disturbance on the state of the spin. When the disturbance introduced from the measurement has negligible effects, these values themselves are termed weak.

\begin{figure}[b]
\centering
{\includegraphics[width= 0.8\columnwidth]{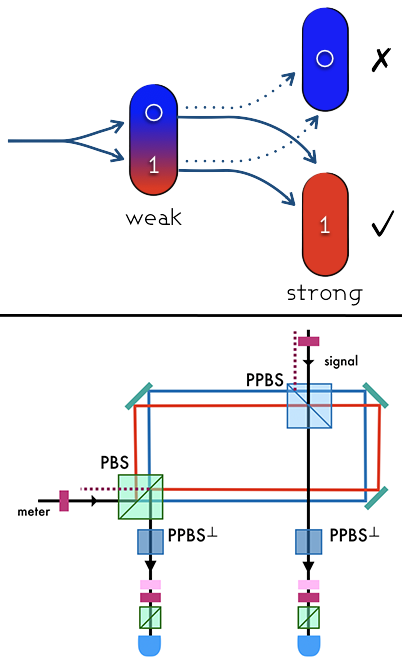}} 
\caption{{\it Upper panel: Conceptual scheme.} A weak measurement is performed on a qubit with possible outcomes 0 and 1, associated to non-orthogonal states. Events are postselected based on a second strong projective measurement associated to a different observable. Coherent effects are known to give rise to different phenomena. {\it Lower panel: Experimental setup.} Signal and meter photons, generated through a Type-I spontaneous parametric down conversion (SPDC) source, are sent through a C-Sign gate \cite{PhysRevLett.95.210504,PhysRevLett.95.210505,PhysRevLett.95.210506,Mancino:2018fk} implemented within a Sagnac interferometer, as shown. This arrangement benefits from the different paths of the H and V polarized beams (red and blue), which allow to decouple the residual signals (purple dashed lines) given by the imperfect partially polarizing beamsplitter (PPBS) transmission with respect to the measured ones.  Two rotated PPBS are used both on the signal and meter to balance polarization losses, and eventually both beams are measured projectively and the coincidence counts are recorded. }
\label{fig:scheme}
\end{figure}
This framework is fully consistent in treating with equal footing pre- and postselection of quantum state, thus making the description more time-symmetric, but some results of this approach seem to make quantum mechanics even more puzzling than it already appears~\cite{Aharonov_2013,Denkmayr:2014rw,PhysRevLett.111.240402}. In some cases, everything can be reconciled with interference effects also taking place for classical waves~\cite{Corr_a_2015}, and it has been suggested that the emergence of anomalous values is an artefact purely due to postselection and also observed in classical probabilities~\cite{PhysRevLett.113.120404}. However, this argument has sparkled a long controversy~\cite{arXiv:1409.5386,arXiv:1409.8555,arXiv:1410.7126,PhysRevLett.114.118901,PhysRevLett.114.118902}. 

Anomalous values, lying outside the spectrum of the observable, are not limited to the weak-value regime, but can emerge at arbitrary disturbance. In the simple case of a single spin-1/2 particle - that nowadays goes more often under the name of qubit - this peculiar effect can be used to flag the failure of a macrorealistic description, as captured by the Leggett-Gard inequality~\cite{PhysRevLett.97.026805,PhysRevLett.100.026804,Palacios-Laloy:2010oz,Goggin1256}; however, this connection does not hold under generic conditions~\cite{PhysRevA.96.052123}. 

This ambiguity prompts the question as to whether generic anomalous values - also outside the weak regime - only bear sense as an accident of quantum interferences in postselection, or if they bring reference to a distinct quantum property.  Recently, it has been shown that anomalous weak values can be directly linked, under specific assumptions, to the fact that no non-contextual theory allowing for ontic states can describe the weak measurement and postselection process~\cite{PhysRevLett.113.200401}. This has been demonstrated in an experiment with single photons~\cite{PhysRevLett.116.180401}. This hints at a general link between anomalies and contextuality.

Remarkably, the notion of contextuality has emerged in yet another aspect of weak measurements: their use for parameter estimation~\cite{Torres:2016zw}. It is a well established result that, since the whole procedure including measurement and postselection can be described as a generalised measurement on the qubit, there is no possibility of using this scheme to improve the precision over standard measurements~\cite{PhysRevLett.114.210801}. The appropriate figure of merit is the Fisher information on the parameter that, in the presence of postselection, can reach higher values than permitted with standard measurements. This is, however, counterbalanced by low postselection probability~\cite{PhysRevA.86.040102,PhysRevLett.112.040406},  bringing the effective Fisher information per prepared event below the standard case. On the other hand, there are cases in which there might exist practical advantages~\cite{Hosten787,PhysRevLett.102.173601,PhysRevLett.115.120401,PhysRevX.4.011031,PhysRevX.4.011032,PhysRevLett.118.070802,Feizpour:2015fk,Dziewior2881}. In Ref~\cite{arXiv:1903.02563}, the usefulness of postselection has been discussed in the light of the overheads of measurements in the cost-benefit budgeting, and, at the same time, put in direct connection with contextuality.

Here we show that for qubits the connection between anomalous values, enhanced Fisher information and contextuality in weak measurements is rather intricate, and no one-to-one connection is possible, other than in the strict weak-value regime. 

{\it Theoretical framework.} The weak measurement we consider is in the form~\cite{PhysRevA.73.012113}:
\begin{equation}
\begin{aligned}
&M_0=\sqrt{\frac{1+\kappa}{2}}\Pi_0+\sqrt{\frac{1-\kappa}{2}}\Pi_1,\\
&M_1=\sqrt{\frac{1+\kappa}{2}}\Pi_1+\sqrt{\frac{1-\kappa}{2}}\Pi_0,
\label{operators}
\end{aligned}
\end{equation}
where $\Pi_x$ is a projector on the state $\ket{x}$ of the qubit, and $\kappa$ is a parameter describing the strength of the measurement, and ranges from $\kappa=0$ for the infinitely unperturbed case, to $\kappa=1$ for the full-strength projective case. Differently from the original setting, operated with a continuous-variable probe with a wide dispersion, we consider the sort of weak measurements obtained by coupling the original qubit to a second probe qubit by means of a quantum logic gate such as a control-NOT~\cite{PhysRevLett.94.220405}. The final postselection on the generic state $\bra{\phi}$ leads to the joint probabilities:
\begin{equation}
p_x=|\bra{\phi}M_x \ket{\psi}|^2,\qquad x=0,1,
\end{equation}
when starting in the state $\ket{\psi}$; these represent the probability of obtaining the outcome $x$, followed by a successful postselection~\cite{PhysRevA.73.012113,PhysRevLett.113.200401}.

Anomalous values are obtained by considering the {\it conditional} probabilities $p^{c}_x=p_x/(p_0+p_1)$; these are hence defined as~\cite{PhysRevA.73.012113,PhysRevLett.94.220405}:
\begin{equation}
\sigma_{\rm w} = \frac{p^{c}_0-p^{c}_1}{\kappa}.
\label{value}
\end{equation} 
Here $\kappa$ serves the purpose of recovering the appropriate scale factor to extract the correct expectation value for the observable $\hat Z=\Pi_0-\Pi_1$ from the weak measurement~\cite{PhysRevLett.92.190402}. This observable has clearly a spectrum ranging from -1 to 1, but anomalous values of $\sigma_{\rm w}$ can be observed for all values $0\leq\kappa<1$~\cite{PhysRevLett.94.220405}.These quantities, however, correspond to the weak values originally introduced in \cite{PhysRevLett.60.1351} only in the limit $\kappa \ll 1$.

\begin{figure}
\centering
{\includegraphics[width= \columnwidth]{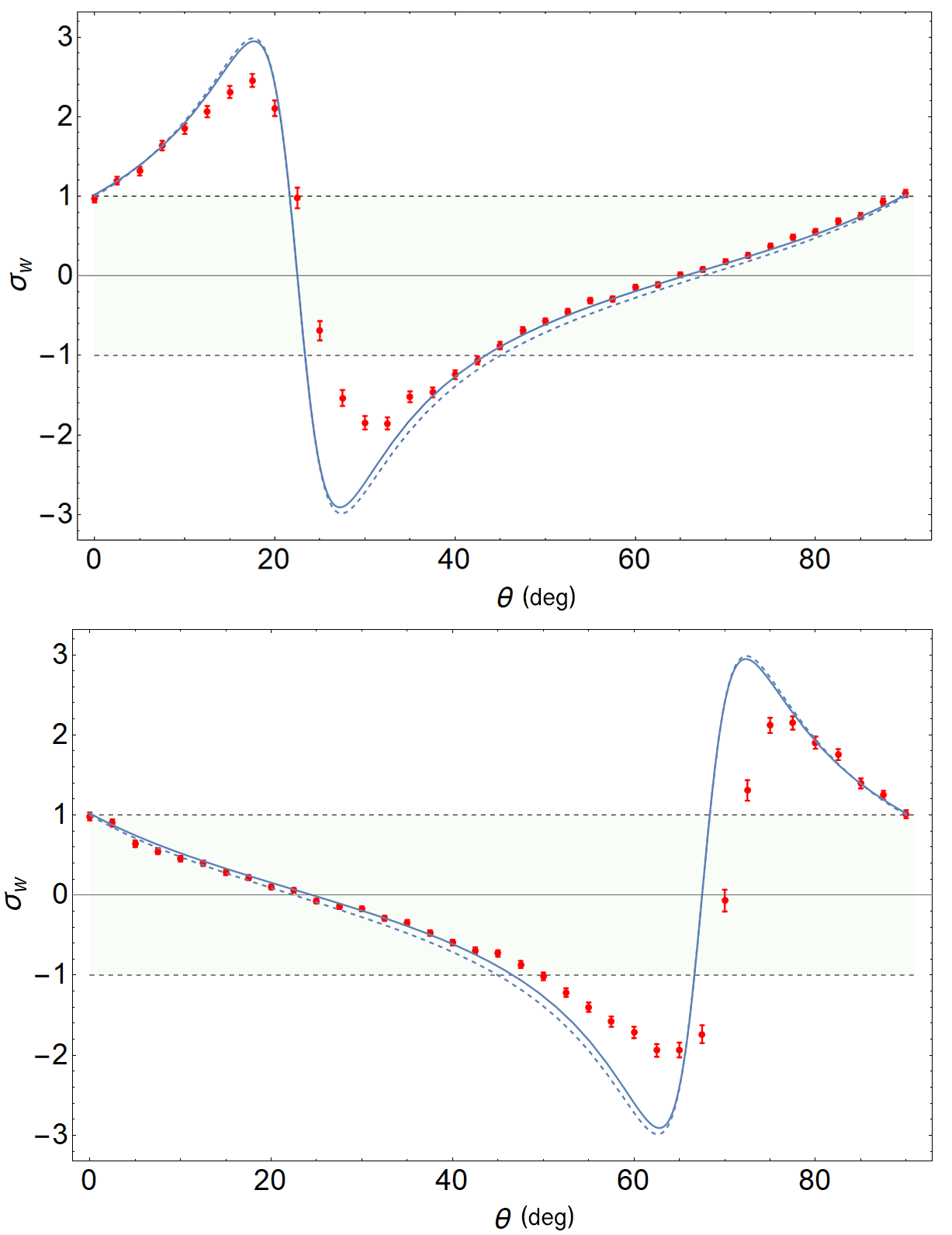}} 
\caption{Measured weak values for both postselections. Each point is determined from coincidence counts accumulated for 5s. The Poissonian statistics of the counts and the uncertainty on $\kappa$ contribute to the error bars. All values outside the shaded region are deemed anomalous. The dashed blue line is the theoretical prediction without experimental imperfections, the continuous blue line is the theoretical prediction taking into account a reduced visibility (v=0.78) and imperfect splitting of the PPBS ($T_H=0.98$ and $T_V=0.34$)}\label{fig:weak}
\end{figure}

If $\ket{\psi}$ contains an unknown parameter $\theta$, the information which can be extracted by means of the measurement and postselection strategy is quantified by the Fisher information (FI)~\cite{doi:10.1142/S0219749909004839}:
\begin{equation}
F_{\rm ps}(\theta)=\frac{(\partial_\theta p^{c}_0)^2}{p^{c}_0}+\frac{(\partial_\theta p^{c}_1)^2}{p^{c}_1}.
\end{equation}
The variance on the parameter $\Delta^2\theta$, obtained from $M_{\rm ps}$ successful postselection events, satisfies the bound $\Delta^2\theta\geq1/(F_{\rm ps}M_{\rm ps})$. The FI derived form the conditional probabilities given the postselection does not obey the same bounds as for standard measurements~\cite{PhysRevA.86.040102}. It is then possible to attain values of $F_{\rm ps}$ higher than the maximal quantum FI $Q$. This does not result in any improvements in the accuracy, since the postselection reduces proportionally the number of repetitions: $F_{\rm ps} M_{\rm ps}\leq Q M_{\rm tot}$, where $M_{\rm tot}$ is the total number of attempts, also including those leading to unsuccessful postselection~\cite{PhysRevLett.112.040406,PhysRevLett.114.210801}. On the other hand, it should be noticed that $F_{\rm ps}>Q$ represents a valid criterion for an anomalous behaviour, since it implies that each repetition carries more FI than the maximum (see also Ref.~\cite{arXiv:1903.02563}). 

Finally, non-contextual models can account for the observed joint probabilities as long as they satisfy the relation~\cite{PhysRevLett.113.200401}:
\begin{equation}
I_0 = \frac{p_0}{p_\phi}-\frac{1+\kappa}{2}-\frac{p_d}{p_\phi}<0,
\label{Pusey}
\end{equation}
where $p_\phi=|\langle\phi,\psi\rangle|^2$, and $p_d=1-\sqrt{1-\kappa^2}$ (see Appendix). A similar inequality can be defined in terms of $p_1$. We remark these are the {\it joint probabilities}, not the conditional ones.

{\it Experiment.} We illustrate in an experiment the connection between the three aspects mentioned above. We prepare qubit states in the form $\ket{\psi}_s=\cos(2\theta)\ket{0}_s+\sin(2\theta)\ket{1}_s$ as superpositions of horizontal and vertical polarisations of a single photon. The measurement operators \eqref{operators} are implemented by coupling the input to the second meter photon in the state $\ket{\mu}=\cos(2\mu)\ket{0}_m+\sin(2\mu)\ket{1}_m$ by means of a Controlled-Sign (C-Sign) gate~\cite{PhysRevLett.94.220405}: this selectively imparts a $\pi$-phase shift to the $\ket{1}_s\ket{1}_m$ contribution with respect to the others. By effect of the coupling, a measurement of the meter qubit in the diagonal polarisation basis, $\ket{\pm}=(\ket{0}\pm\ket{1})/\sqrt{2}$, results in the application of the operators $M_0$ or $M_1$, respectively, with $\kappa=\sin(4\mu)$; in our experiment, the angle $\mu$ has been set to obtain $\kappa=0.335\pm0.008$, as determined following~\cite{PhysRevLett.92.190402}. We further divide our results depending on the postselection event occurred $\bra{\phi}=\bra{-}$ or $\bra{\phi}=\bra{+}$. A summarising scheme is presented in Fig.~\ref{fig:scheme}.

Postselected values $\sigma_w$ are obtained by measuring the coincidence counts associated to $\bra{-}_s\bra{+}_m$ and $\bra{-}_s\bra{-}_m$, from which the conditional probabilities $p_0^c$ and $p_1^c$ can be extracted for different values of $\theta$, Fig.~\ref{fig:weak}; values outside the shaded region are anomalous, as they do not belong to the standard range of expected values. Experimental imperfections, including the visibility limited to $97\%$ of the maximum value on the gate, as well as residual polarisation rotations on both input and meter photons, are responsible for the deviations from the ideal.

The usefulness of this measurement strategy for parameter estimation can be captured as follows:  
The postselected values collected are used to estimate $\theta$, by interpolating the function linking  $\sigma_{\rm w}(\theta)$ to an arbitrary value of $\theta$~\cite{PhysRevA.92.032114}. This has then be used to estimate the uncertainties $\Delta^2\theta$ from those on the postselected values $\Delta^2\sigma_{\rm w}$. The results are reported in Table \ref{table:cr}.

\begin{table}
\begin{center}
\begin{tabular}{ll}
\begin{tabular}{ccc}
$\theta$ $(^{\circ})$ & $\Delta^2\theta$  $(^{\circ})^2$& $\sigma_{CR}$ $(^{\circ})^2$\\
\cline{1-3}
20 & 0,085 & 0,23 \\
\cline{1-3}
22,5 & 0,036 & 0,33 \\
\cline{1-3}
25 & 0,061 & 0,41 \\
\cline{1-3}
27,5 & 0,29& 0,35\\
\end{tabular}
&\quad
\begin{tabular}{ccc}
$\theta$ $(^{\circ})$ & $\Delta^2\theta$  $(^{\circ})^2$& $\sigma_{CR}$ $(^{\circ})^2$\\
\cline{1-3}
67,5 & 0,12 & 0,37 \\
\cline{1-3}
70 & 0,041 & 0,43 \\
\cline{1-3}
72,5 & 0,079 & 0,43 \\
\cline{1-3}
75 & 0,76 & 0,3\\
\end{tabular}
\end{tabular}
\end{center}
\caption{Comparison with Cramér Rao variance. Left: comparison between the measured variance and the Cramér-Rao one for the $\langle\phi\vert = \langle - \vert$ postselection. Right: comparison between the measured variance and the Cramér-Rao one for the $\langle\phi\vert = \langle +\vert$ postselection.}
\label{table:cr}
\end{table}

Differently from the estimation of the interaction strength in~\cite{PhysRevA.86.040102}, where the postselected values approximately quantify the Fisher information on the parameter of interest, here the connection between anomalies in the observed value of $\sigma_{\rm w}$ and an improved Fisher information $F_{\rm ps}$ is tenuous. The reason is in the fact that, writing the probabilities $p^c_0$, $p^c_1$ as a function of $\sigma_{\rm w}$, as expressed in \eqref{value}, the Fisher information takes the form:
\begin{equation}
F_{\rm ps}=\frac{\kappa^2\left(\partial_\theta\sigma_{\rm w} \right)^2}{1-\kappa^2\sigma_{\rm w}^2},
\end{equation}
hence it is expressed only in terms of the quantity $\kappa\sigma_w$, which attains values between -1 and 1. The origin of the improved Fisher information is then not associated to the mere presence of anomalous values, but to the fact that postselection leads to a change in the functional shape of $\sigma_{\rm w}(\theta)$ with respect to the standard case. This change has been proven useful for superior alignment procedures~\cite{Dziewior2881}.

A different interpretation can be drawn in terms of non-Hermitian operators~\cite{PhysRevLett.80.5243}. Simple calculations show that the full measurement process, including the weak ones  \eqref{operators}, and the projective measurement in the $\vert\pm\rangle$ basis leading to postselection, is equivalent to a four-outcome generalised measurement. Each outcome is associated to a Bloch vector at the angles $\pi/2\pm4\mu$, and $-\pi/2\pm4\mu$ with respect to the $Z$ axis, corresponding to one of the possible coincidence events. Postselection amounts to ignoring the first two outcomes, as if the measured operators were PT-symmetric non-Hermitian, in complete analogy with the behaviour observed in~\cite{arXiv:1812.05226,arXiv:1901.07968}.

We now turn our attention to the connection between the emergence of anomalous values and a possible relation to contextuality. The existing link between anomalous weak values and the Leggett-Garg inequality, holding for any measurement strength \cite{PhysRevLett.100.026804} may lead into thinking of a similar resilient connection to contextuality.

\begin{figure}
\centering
{\includegraphics[width= \columnwidth]{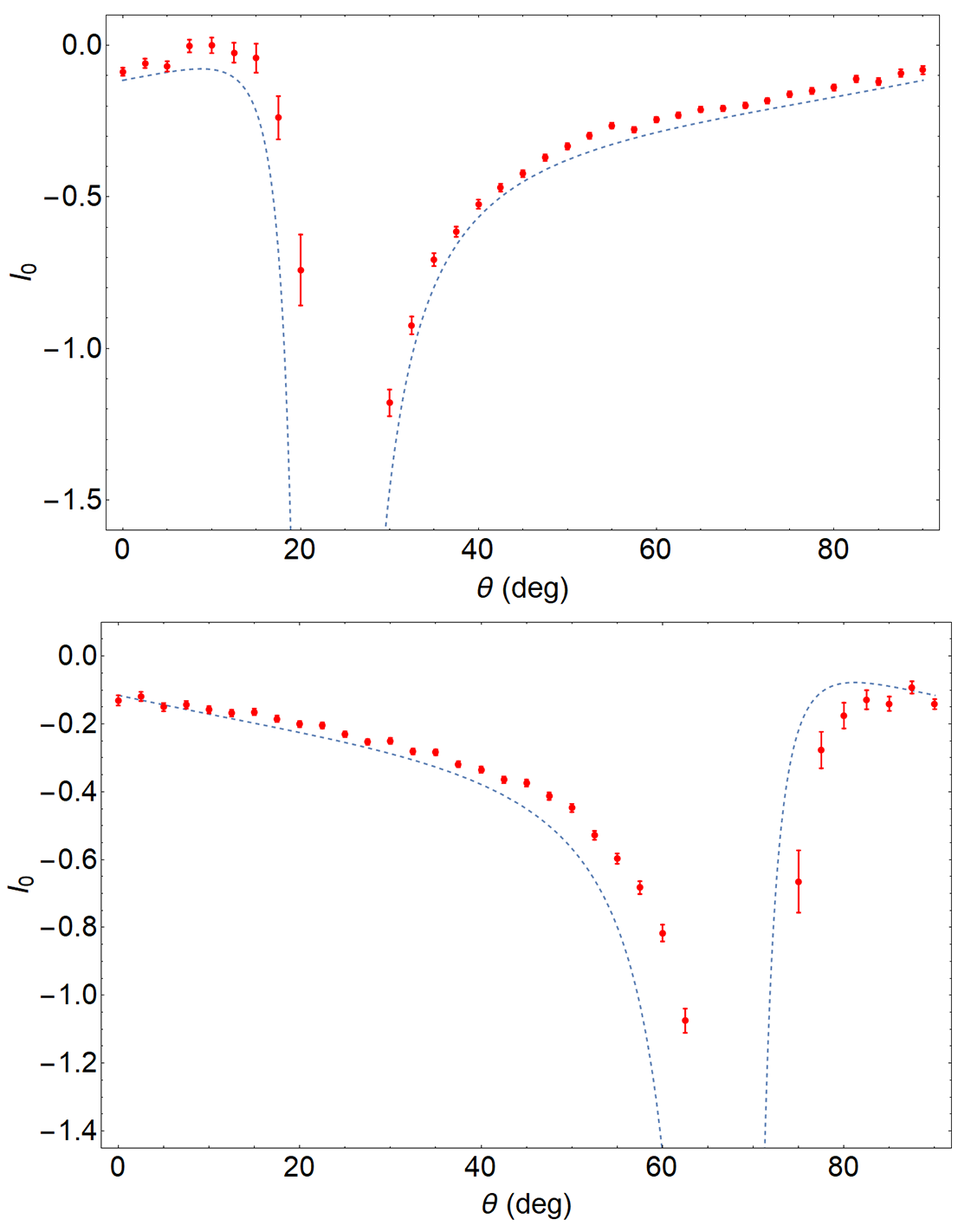}} 
\caption{Verification of Pusey inequality for both postselections. The left hand side of inequality (5) is calculated under the best circumstances for disproving the non-contextual model. }
\label{fig:pusey}
\end{figure}

The data in Figure \ref{fig:pusey} are obtained by evaluating the probabilities in Eq \eqref{Pusey} from the experimental coincidence counts. Despite the fact we register anomalous values for a considerable part of the inputs, none of these are able to demonstrate the unsuitability of a non-contextual model.  

This result was in part foreshadowed in the criticism raised by \cite{arXiv:1410.7126} about the classical mechanism to obtain anomalous values with classical probabilities in \cite{PhysRevLett.113.120404}. The argument against the classical explanation, highlighted the negligible disturbance imposed on the state as the key point. As our experiment departs from those conditions, a non-contextual model becomes appropriate. The identification of the concepts of anomalous value and non-contextuality is legitimate only in the proper weak limit of negligible measurement strength. In all other cases anomalies might provide an indication, but then they have to be more pondering.  Inequality in Eq. \eqref{Pusey} provides the quantitative means for such an analysis, although admittedly does not exhaust all possible scenarios. 

Finally, a comment is opportune between the improved Fisher information and contextuality, whose connection has been identified in \cite{arXiv:1903.02563}. Our results do not contradict this claim because our postselection scheme implements measurement and postselection in the opposite order than the one in \cite{arXiv:1903.02563}.  This restriction ultimately results from working with qubits: postselecting before the measurement would erase all information about the relevant parameter. Postselection is futile, the measurement should be terminated.  Contextuality then provides useful schemes in postselected metrology, however once again there is no complete overlap between the two concepts. 

\section*{Conclusions} 

In this article we have shown how a simple case of a qubit illustrates subtleties in the relation among three different aspects related to a measurement followed by a postselection. The take home message is that results holding valid in the weak regime cannot be exteded when the disturbance becomes sizable. We envisage that these results, albeit partially negative, may stimulate more refined approaches to comprehend the quantumness of this scheme.


%

\section*{Acknowledgements}
The authors would like to thank M. Genovese for fruitful discussion. 

\section*{Appendix} We extend Pusey's original proof in Ref.~\cite{PhysRevLett.113.200401}, tailored to meter states with a continuous spectrum, to encompass measurement operators in the form \eqref{operators}. The aim is to find quantitative bounds to which theories admitting ontic states $\lambda$ obey. For the sake of simplicity, we will adopt the terminology of quantum mechanics, though all the statements can be modified in order not to make explicit reference to it~\cite{PhysRevLett.116.180401}.

The response of the prepared state $\ket{\psi}$ to any positive-operator valued measurement (POVM) can then be rewritten as~\cite{PhysRevLett.113.200401}:
\begin{equation}
\bra{\psi}E_x\ket{\psi}=\int_\Lambda p(E_x|\lambda)p(\lambda)d\lambda,
\end{equation}
where $p(\lambda)$ is the distribution of the ontic states corresponding to the preparation $\ket{\psi}$, and $p(E_k|\lambda)$ is a conditional probability, which, in a non-contextual model, can not depend on how the POVM is actually implemented. Further, we assume that, when $E_k=\ket{\phi}\bra{\phi}$ is a projector, $p(\ket{\phi}\bra{\phi}|\lambda)$ is either 0 or 1, i.e. sharp measurements have deterministic outcomes. The overlap $p_\phi = |\langle\phi|\psi\rangle|^2$ is then calculated as the ensemble average $\int_\Lambda p(\ket{\phi}\bra{\phi}|\lambda) d\lambda$.

Our measurement corresponds to the POVM elements
\begin{equation}
\begin{aligned}
&E_0=\frac{1+\kappa}{2}\Pi_0+\frac{1-\kappa}{2}\Pi_1,\\
&E_1=\frac{1+\kappa}{2}\Pi_1+\frac{1-\kappa}{2}\Pi_0.
\label{POVM}
\end{aligned}
\end{equation}
If the outcomes of the weak measurement is ignored, postselection on the state $\bra{\phi}$ can be described by the operator
\begin{equation}
S = M_0\ket{\phi}\bra{\phi}M_0^\dag+M_1\ket{\phi}\bra{\phi}M_1^\dag.
\end{equation}
This can be decomposed as
\begin{equation}
S = (1-p_d)\ket{\phi}\bra{\phi}+p_d E_d,
\label{eqnS}
\end{equation}
for some POVM with elements $E_d$, and $I-E_d$. Explicit calculations show that $p_d =1-2\sqrt{1-\kappa^2}$. Finally, we consider the cases for which $p_\phi>0$.

Under these circumstances, one can demonstrate that non-contextual models imply the inequality \eqref{Pusey}; the proof follows closely the original one due to Pusey \cite{PhysRevLett.113.200401}, with a minor modification to account for the explicit form of the POVM elements \eqref{POVM}. We start by decomposing the measurement operator $E_x$ as a ``consolidated  measurement", constituted by $S_x=M_x\ket{\phi}\bra{\phi}M_x^\dag$, and $F_x=M_x(I-\ket{\phi}\bra{\phi})M_x^\dag$. The element $S_x$ fulfills the relation:
\begin{equation}
|\bra{\phi}M_x\ket{\psi}|^2=\bra{\psi}S_x\ket{\psi}= \int_\Lambda p(S_x|\lambda)p(\lambda)d\lambda. 
\end{equation}
The conditional probability for $E_0$ admits two equivalent decompositions by virtue of non-contextuality:
\begin{equation}
\begin{aligned}
p(E_0|\lambda)=&p(S_0|\lambda)+p(F_0|\lambda)\\
=&\frac{1+\kappa}{2}p(\Pi_0|\lambda)+\frac{1-\kappa}{2}p(\Pi_1|\lambda).
\end{aligned}
\end{equation}
We then obtain the chain of inequalities
\begin{equation}
p(S_0|\lambda)\leq p(E_0|\lambda) \leq \frac{1+\kappa}{2}.
\end{equation}
We turn our attention to the decomposition \eqref{eqnS} that, invoking non-contextuality in a similar fashion as above, gives: 
\begin{equation}
\begin{aligned}
p(S|\lambda)=& p(S_0|\lambda)+ p(S_1|\lambda)\\
=&(1-p_d)p(\ket{\phi}\bra{\phi}|\lambda)+p_d p(E_d|\lambda).
\end{aligned}
\end{equation}
The set $\Lambda$ of the ontic states can be separated in the kernel $\Lambda_0$ of $p(\ket{\phi}\bra{\phi}|\lambda)$, and its complement $\Lambda_1$; on the former subset, we have 
\begin{equation}
p(S_0|\lambda)\leq p(S|\lambda) \leq p_d.
\end{equation}
Therefore, we can write a second chain of inequalities
\begin{equation}
\begin{aligned}
|\bra{\phi}M_0\ket{\psi}|^2&\leq\int_{\Lambda_1} p(S_0|\lambda)p(\lambda)d\lambda+p_d\\
&\leq \frac{1+\kappa}{2} \int_\Lambda p(\ket{\phi}\bra{\phi}|\lambda)p(\lambda)d\lambda+p_d\\
&\leq \frac{1+\kappa}{2} p_\phi+p_d,
\end{aligned}
\end{equation}
hence proving our statement.

\bibliography{BiblioWeakValues}

\end{document}